\documentclass[journal]{vgtc}                     

\onlineid{1485}

\vgtccategory{Research}

\vgtcpapertype{please specify}

\title{``May I Speak?": Multi-modal Attention Guidance in \\Social VR Group Conversations}

\author{%
  \authororcid{Geonsun Lee}{0000-0001-9401-8559},
  Dae Yeol Lee, Guan-Ming Su, and 
  Dinesh Manocha
}

\authorfooter{
  \item
  	Geonsun Lee is with University of Maryland.
  	E-mail: gsunlee@cs.umd.edu.
  \item
  	Dae Yeol Lee is with Dolby Laboratories.
  	E-mail: DaeYeol.Lee@dolby.com.

  \item Guan-Ming Su is with Dolby Laboratories.
  	E-mail: guanming.su@dolby.com.

   \item Dinesh Manocha is with University of Maryland.
  	E-mail: dmanocha@umd.edu.
}

\abstract{%
  In this paper, we present a novel multi-modal attention guidance method designed to address the challenges of turn-taking dynamics in meetings and enhance group conversations within virtual reality (VR) environments. Recognizing the difficulties posed by a confined field of view and the absence of detailed gesture tracking in VR, our proposed method aims to mitigate the challenges of noticing new speakers attempting to join the conversation. This approach tailors attention guidance, providing a nuanced experience for highly engaged participants while offering subtler cues for those less engaged, thereby enriching the overall meeting dynamics. Through group interview studies, we gathered insights to guide our design, resulting in a prototype that employs ``light" as a diegetic guidance mechanism, complemented by spatial audio. The combination creates an intuitive and immersive meeting environment, effectively directing users' attention to new speakers. An evaluation study, comparing our method to state-of-the-art attention guidance approaches, demonstrated significantly faster response times ($p < 0.001$), heightened perceived conversation satisfaction ($p < 0.001$), and preference ($p < 0.001$) for our method. Our findings contribute to the understanding of design implications for VR social attention guidance, opening avenues for future research and development.
}

\keywords{Social VR, Attention Guidance, Multi-modal Interaction, Group Conversations}

\teaser{
  \centering \includegraphics[width=1.0\textwidth]{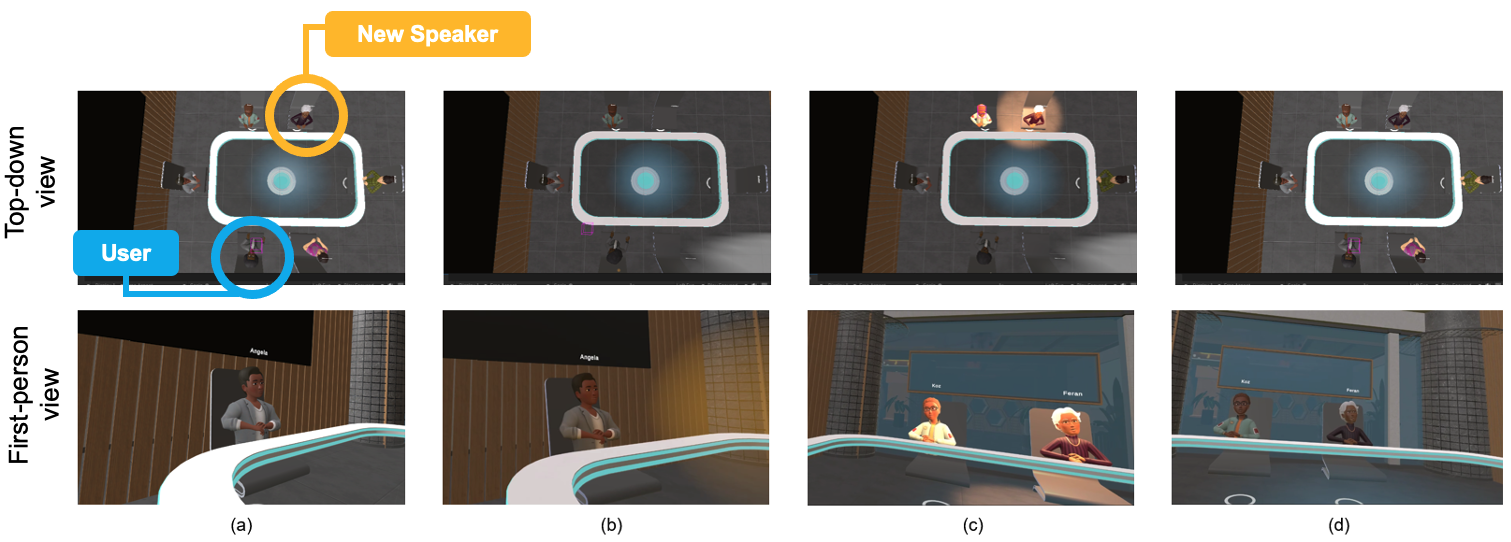}
  \vspace{-1em}
  \caption{We present a diegetic multi-modal attention guidance method designed for group conversations in social VR. The top row illustrates a top-down view of a virtual conference room, while the bottom row presents the first-person view. When (a) an agent (target) positioned out of the user's view signals a desire to speak; (b) The environment light dims, a point light affixed to the user's right periphery of the screen instantiates, and a spatial sound effect with a chiming sound commences; (c) As the user turns their head towards the target, the environment light brightens, the point light vanishes, a spotlight illuminates the target, and the chiming sound emanates from the target's location; (d) Once the user's gaze aligns with the target, the spotlight and chiming sound dissipate. Our approach demonstrated statistically significant enhancements in users' satisfaction with the conversation, faster response times, and higher preference compared to conventional methods.
    }
}

\graphicspath{{figs/}{figures/}{pictures/}{images/}{./}} 

\usepackage{tabu}                      
\usepackage{booktabs}                  
\usepackage{lipsum}                    
\usepackage{mwe}                       

\usepackage{mathptmx}                  

\usepackage{amsmath}
\usepackage{tabularx}
\usepackage{adjustbox}
\usepackage{appendix}
\usepackage{hhline}

\begin{document}



\maketitle

\section{Introduction}
\label{sec:intro}
Social virtual reality has gained traction as a compelling solution for enabling people to interact within shared virtual environments with head-mounted displays (HMDs). This field encompasses an array of commercially available applications, including VRChat, Spatial, RecRoom, Mozilla Hubs, Glue, Horizon Worlds, and others. While initially geared towards casual socializing, these platforms have progressively evolved to cater to diverse use cases, extending into professional domains such as conferences and business-focused meetings~\cite{rubinger2020maximizing}. The ability of VR to provide an immersive and shared spatial experience adds a layer of depth to interactions, setting the stage for a paradigm shift in remote collaboration and communication.

In the context of remote communication, the issue of {\em turn-taking}  has emerged as a persistent challenge in effectively supporting remote meetings. This challenge, which has been the subject of scholarly inquiry for decades~\cite{sellen1995remote, vertegaal1999gaze}, has become especially pronounced during the widespread adoption of remote work necessitated by the COVID-19 pandemic~\cite{rintel2020study}. The intricacies of managing turn-taking in the context of video-mediated communication have become a focal point, with numerous contributing factors~\cite{buxton2009mediaspace,finn1997video,nash2021nomadic}. The inherent difficulty in discerning non-verbal cues, such as gestures and head/body movements, leads to disruptions and inefficiencies in the communication flow.

While VR offers a promising alternative by simulating face-to-face meetings and providing cues based on spatial layout and proximity~\cite{zhang2002social,xu2017attention,williamson2022digital}, it needs to deal with many other challenges. The limitations of immersive VR, such as a restricted field of view and imperfections in face and gesture tracking, pose unique obstacles that impact the effective identification of new speakers in a conversation. Additionally, individuals with conditions such as autism or those with reduced social skills may face difficulties participating in social group conversations~\cite{girolametto1988improving, stanton2011promoting}. Despite VR's potential to address these challenges by serving as a training ground for social skill development~\cite{jarrold2013social,adjorlu2018head} and enabling novel interactions that are beyond simply imitating reality~\cite{hollan1992beyond,sabri2023challenges}, there remains an unmet need for comprehensive solutions that facilitate natural signaling of the desire to speak and enhance the perceptibility of such signals by other users.

\noindent {\bf Main Results:} In this paper, we present improved methods related to improving user experiences related to VR turn-taking. Our goal is to facilitate the recognition of new speakers in a mid-sized group conversation, thereby guiding users' attention effectively. Our approach uses attention guidance methods,  which have been extensively studied in the context of VR, and seek to direct users' attention with rapid responsiveness while ensuring minimal disruption to their immersion~\cite{lange2020hivefive}. To this end, we introduce a diegetic multi-modal attention guidance approach that utilizes both lights and spatial audio within the virtual meeting environment. Our approach is designed to enhance the user experience by enabling the identification of new speakers, all while maintaining the seamless flow of ongoing conversations and sustaining immersion within the VR environment. Additionally, our formulation strives to enhance social presence within these virtual interactions.

Our methodology involves a group interview study with experienced users in VR meeting platforms to elucidate key design considerations. Based on these considerations, we propose an approach where attention guidance dynamically adjusts based on a user's engagement in an ongoing conversation. We present various components of our approach and illustrate their interactions within the virtual environment.

Subsequently, we conduct a user study simulating participants engaging in conversations within a social VR setting, employing avatars with pre-recorded interactions. We perform an evaluation corresponding to when a new virtual speaker signals their desire to speak in two scenarios: (i) while the user is already engaged in conversation and (ii) while the user is in listening mode. We compare our approach against two existing methods: a text window and icon indication used in the VR meeting application Horizon Workrooms, and Subtle Gaze Direction (SGD)~\cite{grogorick2017subtle}.
Our evaluation results demonstrate the effectiveness of our method in guiding users' attention, positively impacting perceived conversation quality, and high preference. Our novel contributions include:
\begin{itemize}
    \setlength\itemsep{-0.5em}
    \item Insights into the design of attention guidance methods for social VR derived from a comprehensive study involving user group interviews with experienced participants.
    \item The novel concept of an engagement-based attention guidance approach for turn-taking in social VR group conversations.
    \item A prototype of a diegetic multi-modal attention guidance method utilizing light and spatial audio within the virtual environment.
    \item Highlighting the significant impact of our multi-modal attention guidance approach on response time, perceived conversation quality, and achieving the highest overall preference when compared to traditional methods.
    . 

\end{itemize}

\section{Related Work}
\label{sec:related_work}
\subsection{Social VR}
Social virtual reality (Social VR) has recently gained considerable attention from the human-computer interaction (HCI) and VR communities~\cite{j2022social} exploring spatial navigation and social mechanics~\cite{mcveigh2018s, moustafa2018longitudinal}. A longitudinal study by Moustafa et al.~\cite{moustafa2018longitudinal} revealed the transferability of existing social group dynamics to VR interactions. Other studies have delved into social interactions on Social VR platforms, addressing challenges such as mitigating harmful behavior~\cite{j2022social, mcveigh2019shaping}.

Prior studies have demonstrated that self-embodiment and non-verbal cues play pivotal roles in establishing social presence within social VR applications~\cite{yassien2020design}. To enhance user experiences in social interactions, endeavors have been made to amplify these non-verbal cues~\cite{freeman2021hugging}. Gaze direction emerges as a prominent method for social signaling, facilitating the seamless transition of users from the awareness of others' presence to interactive engagement. Consequently, researchers have been actively engaged in the development of gaze-tracking VR solutions~\cite{babu2017gaze, yassien2020design}. Roth et al.'s work~\cite{roth2018beyond} involved augmenting social behaviors within a multi-user virtual museum environment. Their experimentation included visualized eye contact represented by bubbles, highlighted joint attention through visual cues, and color-coded group affiliations.

\subsection{Social Attention in Group Conversations}
Turn-taking describes the dynamic flow of participation among speakers in a conversation over time. Conversations involve a constant reshuffling of participation, and it is defined that participation shifts occur in four different types: {\em turn-receiving, turn-claiming, turn-usurping, and turn-continuing}~\cite{gibson2003participation}. When a person speaks after being addressed, it is termed turn-receiving; if a person speaks after someone else is addressed, it is called turn-usurping. When a person speaks after someone addresses the group, it is termed turn-claiming. Finally, when someone who is already talking changes targets, it is termed turn-continuing. While turn-taking in conversation is often unconstrained and unplanned, in institutional settings, the system has been described as more restricted and specialized~\cite{heritage1992talk}.

Common indicators of the desire to speak in a group include raising hands and nodding heads. In some settings, participants need to be selected by a moderator and queued to be the next speaker~\cite{mondada2013embodied}. In face-to-face conversations, the next speaker is usually indicated by social attention or eye contact~\cite{langton2000eyes, ho2015speaking, maran2021visual}. Dawson and Foulsham~\cite{dawson2022your} investigated how shifts in attention between speakers depend on visual or auditory cues. They found that eye-tracked participants often fixated on the person speaking and shifted their gaze in response to changes in the speaker, even when sound was removed or the video was freeze-framed.

The underlying medium used also influences conversation patterns and social attention~\cite{abdullah2021videoconference}. In video conferencing, users can use the `raise-hand' feature to signal their intention to speak and wait to be addressed by others in the group~\cite{rintel2020study}. Hu et al.~\cite{hu2023openmic} developed OpenMic, an interface that visualizes conversational floor transitions by incorporating proxemic metaphors in a videoconferencing system. Steptoe et al.~\cite{steptoe2008eye} introduced one of the first avatar gaze tracking systems and provided preliminary evidence that it improves communication. In VR, Li et al.~\cite{li2022conversation} introduced a shared VR environment visualization to aid in conversational turn-taking, employing cylinders that expand over time to represent the duration of each speaker's turn, thereby facilitating balanced participation in the conversation.

Within the realm of turn-taking, our approach focuses on the participation shifts of turn-claiming and turn-usurping. This occurs when users are perceived to primarily use the 'raise-hand' feature or when the person who wants to speak needs to grab the social attention of the group because they are not directly addressed to speak.



\subsection{Guiding User's Attention}
The exploration of guiding user attention spans across diverse mediums, from the realm of 2D images to the immersive experiences of VR and AR~\cite{rothe2019guidance}. The aim is to steer users toward specific focal points intended by creators. In the VR community, users often confront the challenge of significant scene details lying beyond their field of view, instigating a concern of missing out on crucial elements~\cite{tse2017there,macquarrie2017cinematic}.

A technique to address this involves direct modifications to images. Smith et al.~\cite{smith2013nonblurred} employed a combination of blurred and non-blurred areas in videos, directing viewers towards regions with minimal spatial blur when the rest of the image is intentionally blurred. Hata et al.~\cite{hata2016visual} explored thresholds where blur effects can be subtly applied, effectively guiding users visually. Stylistic rendering, encompassing control over depth of field, colors, brightness, and sharpness, represents another avenue. Additionally, the simulation of brightness contrast through lights, a technique well-established in film, has been leveraged for effective visual attention~\cite{cole2006directing}. El et al.~\cite{el2009dynamic} implemented ALVA (Adaptive Lighting for Visual Attention), dynamically adjusting lighting color and brightness to enhance visual attention within gaming environments.

A prevalent category in attention guidance involves diegetic methods~\cite{nielsen2016missing,sheikh2016directing,rothe2018guiding}, where cues seamlessly integrate into the scene, encompassing elements like characters, lights, or sounds. This concept, rooted in film theory as diegesis, entails elements belonging to the narrative world.  Diegetic visual cues, such as a person looking in a certain direction, significantly influence the viewer's gaze, as observed with moving objects~\cite{sheikh2016directing}. Auditory cues within the diegetic framework leverage sound to prompt users to search for the source, prompting a change in their viewing direction~\cite{rothe2018guiding}. Noteworthy for their subtlety, diegetic cues afford viewers the freedom to follow them naturally. More recently, Lange et al.~\cite{lange2020hivefive} introduced Hive Five, a particle effect emulating the swarm motion of bees as a cursor, showcasing its ability to guide users' attention while preserving immersion.

Various techniques have been explored to accentuate crucial content in VR, ranging from methods utilizing different shapes like arrows~\cite{chittaro20043d,lin2017tell} to halos~\cite{gruenefeld2018beyond}. Subtle Gaze Direction (SGD), leveraging eye tracking, subtly guides users' gaze without their explicit awareness~\cite{bailey2009subtle}. This method modulates a target region in the peripheral area, encouraging viewers to direct their gaze while discontinuing modulation when the viewing direction aligns to a certain degree. Luminance modulation and warm-cool modulation were developed, employing flickering at 10 Hz within a circular region. Grogorick et al.~\cite{grogorick2017subtle} later adapted this method for VR environments.

Multi-modal attention guidance represents an advanced approach. Reyes et al.~\cite{reyes2021multi} developed a guiding method integrating both visual and auditory cues, with their study demonstrating the superior efficacy of incorporating an auditory cue alongside visual cues.

While a myriad of methods exists, none have systematically addressed their adaptation to social meeting settings, characterized by pre-existing social attention dynamics and diverse individual social engagements. Our method takes these nuanced considerations into account, dynamically adjusting the intensity and components of the attention guidance framework presented to the user.


\subsection{Notifications in VR/AR}
A notification is a proactive delivery of information to users through visual, auditory, or haptic alerts designed to attract their attention~\cite{iqbal2010oasis,pielot2018dismissed}. Ghosh et al.~\cite{ghosh2018notifivr} investigated notifications in VR, comparing visual, audio, and haptic modalities, as well as their pairwise combinations. The study found that both audio and haptic modalities effectively elicited reactions from participants in VR. However, haptic notifications faced challenges such as confusion or being missed due to interactions with existing objects in the VR environment. In a related study, George et al.~\cite{george2018intelligent} conducted an exploratory lab study comparing three notification types: text, spotlight, and global light. Text notifications prompted quick responses but exhibited the lowest presence, while ambient light showed the lowest attention-grabbing but the highest presence. This trade-off highlights the importance of considering both aspects in notification design. To minimize disruption, Chen et al.~\cite{chen2022predicting} identified opportune times for delivering notifications in VR, allowing for their scheduled presentation. Rzayev et al.~\cite{rzayev2019notification} investigated efficient notification presentation in VR by comparing different placements of notifications in various tasks. They also explored the position of notifications in AR glasses and how they would be perceived during face-to-face communication~\cite{rzayev2020effects}.

It's noteworthy that existing works on notifications in VR primarily focus on conveying information from the external world to users who are obscured by wearing a VR HMD. Similarly, our work aligns with the concept of notifications as we aim to convey information to users engaged in tasks that are susceptible to "interruption." Users may revisit a signal indicating a new speaker's intention to speak, similar to the way notifications are revisited after initially noticing them once the current speaker is done.


\section{Formative Study: Group Interview}
\label{sec:formative_study}
Our primary goal was to precisely identify the issue at hand and develop a corresponding interaction strategy. We conducted a group interview study, focusing on two main objectives: firstly, to pinpoint the gap between user needs for turn-taking in social VR and current attention guidance methods; and secondly, to determine the desired features for attention guidance in this context. This study involved seven experienced VR users, all actively participating in VR meetings.

\subsection{Participants}
For the group interview, we conducted a pre-screening process and selected participants with substantial VR application experience and engagement in regular or irregular VR meetings. Seven participants were recruited, each with 2 to 10 years of VR experience and over a year of professional VR meeting attendance. The experienced meeting sizes ranged from small (up to 5 attendees) to medium-sized groups (5 to 14 attendees). Participants had prior exposure to VR platforms such as Spatial (5 participants) and Horizon Workrooms (3 participants), along with others like Glue or BigScreen VR. All had access to an Oculus Quest 2 or Oculus Quest Pro.

\subsection{Study Setup and Procedure}

\subsubsection{Prototype of Existing Attention Guidance Methods}
We captured videos of prototypes featuring various existing attention guidance methods within a VR meeting setting. Four methods from previous work were implemented: arrow, SGD~\cite{bailey2009subtle}, Hive Five~\cite{lange2020hivefive}, and conventional text-based notifications akin to video conference platforms. We use the Unity game engine for developing the prototype, with each video lasting around 15 seconds. These videos showcased the user looking at a virtual avatar, the application of the attention guidance method, and the user's view being directed to an avatar on the far right. 
Due to computational constraints faced by some participants, we presented existing methods using videos, ensuring uniform exposure and circumventing technical limitations. This approach facilitated immediate, collaborative discussion and analysis of specific method aspects with visual aids.

\subsubsection{Procedure}
The group interview was conducted in two sessions to accommodate participant schedules, with three and four participants per session. The interviews took place remotely via Meta Horizon Workrooms to introduce newcomers to Horizon Workrooms' features and enable instant demonstration of comments and ideas within VR. The Workrooms' layout was set to ``Meeting." Participants were given a 5- to 10-minute platform orientation before the actual interview to prevent distractions during the session. The interview comprised three phases: (1) Initial questions focused on general VR meeting experiences, benefits and limitations compared to in-person and video calls, methods of grabbing attention to speak, instances of cues being missed, and hand-raising behavior; (2) Prototype videos of existing attention guidance methods were presented using screen share within Horizon Workrooms. This was followed by a try-out of Horizon Workrooms' ``raise hand" feature, which participants explored within different seat positions for 3 to 5 minutes; (3) Participants shared opinions on the methods from phase two, suggested additional features, and brainstormed. The study duration was approximately 75 minutes.

\subsection{Findings}
\paragraph{Positive Aspects of VR Meetings}
Participants in the interview highlighted intuitive interactions, like head-turning to see others and discerning directional sounds, which enhance the "social presence" in VR meetings, setting them distinctly apart from traditional video calls. They also noted the benefits of customizable environments tailored to attendee count and meeting type, and an increased focus enforced by wearing VR headsets, offering clear advantages over in-person meetings.

\paragraph{Limitations in VR Meetings:}
Despite their strengths, current VR-based meetings have many limitations. Foremost among these is the limited field of view in VR headsets, which impairs peripheral awareness and the ability to notice distant users and their nonverbal cues for speaking. While recent advancements have improved body and facial tracking, participants noted a disparity between real-world gestures and their VR counterparts, resulting in less expressive and sometimes ambiguous non-verbal cues. Technical issues like network lag and tracking errors disrupt the fluidity of conversation, causing significant inconvenience.

\paragraph{Social Attention in VR Meetings:}
Drawing parallels to in-person meetings, participants signaled their intent to speak in similar ways: i) emitting sounds like throat clearing or table tapping, and ii) using gestures such as head turns, nods, or virtual hand raises. Notably, participants highlighted scenarios where such cues are missed. Instances included complex discussions that deter opportune contributions, participants positioned out of others' views while concentrating on someone else, and challenges in noticing others within larger groups. These turn-taking issues were particularly prominent in formal meetings where participants are less acquainted and no dedicated moderator exists.

\paragraph{Feedback on Existing Attention Guidance Methods:}
Participant feedback on the presented attention guidance methods within a social context was strikingly uniform. A consensus emerged among participants that the demonstrated methods proved overly distracting. The utilization of screen-fixed UI elements like arrows or text-based notifications, despite their intention to guide attention, was deemed overly intrusive by most. Among the four methods, SGD was acknowledged as the most subtle, yet its constant flickering in the user's peripheral vision was deemed distracting. One participant likened the SGD's flickering effect to an ``alert" rather than a guidance mechanism, attributing this to its design for on-screen visual targets rather than off-screen objects as is common in social VR. HiveFive's swarm motion was considered distracting and out of sync with the meeting room environment, undermining its usability. Participants collectively expressed that existing methods prioritize directing attention to a target, overlooking the nuances of group conversations. 

As for the ``raise hand” feature in Horizon Workrooms, participants acknowledged its usefulness mainly because they found it similar to those in video-conference platforms like Zoom or Microsoft Teams. However, they noted its potential to be overlooked, especially in larger groups or by those lacking social skills to naturally draw attention.

\paragraph{Desired Characteristics in Social Attention Guidance:}
Participants' brainstorming suggestions revealed recurring themes aligning with their preferred attention guidance characteristics, emphasizing context-dependent effectiveness. For example, informal gatherings might not require guidance, in contrast to formal or large meetings without a clear moderator. After transcribing and coding these inputs, we identified five key characteristics, informed by prior discussions on VR meetings and existing attention guidance methods.

\begin{itemize}
    \setlength\itemsep{-0.5em}
    \item \textbf{Diegetic:} Participants advocated for attention guidance cues that seamlessly integrate into the environment, avoiding excessive user distraction.
    \item \textbf{World-Referenced:} Echoing the ``diegetic" principle, participants preferred cues that are fixed to the virtual world's elements rather than to the screen.
    \item \textbf{Subtlety:} Desired descriptors for the method included ``subtle," ``ignorable," and minimally disruptive. Interviewees stressed this quality's importance for speakers, listeners, and those interjecting, with a balance between unobtrusiveness and efficient signaling seen as crucial.
    \item \textbf{Control of Urgency:} Four participants articulated a desire for nuanced control over the degree of attention they receive. They envisioned situations where they might prefer to go unnoticed or, conversely, urgently seek recognition based on the content of their contribution. This nuanced approach contrasts with existing ``raise hand" features that typically represent user intention in binary terms.
    \item \textbf{Multi-Modality:} Participants consistently highlighted the significance of audio cues. One participant underscored that VR meetings stand out due to their spatial audio dimension, which aids users in identifying sound direction and speaker location. Participants proposed augmenting this auditory spatial awareness with corresponding visual cues, such as modulating a speaker's volume or introducing non-intrusive chime sounds from the new speaker's direction. Vocal narration was generally discouraged due to its potential to disrupt the conversation.
\end{itemize}

\begin{figure*}
	\centering
	\includegraphics[width=1.0\textwidth]{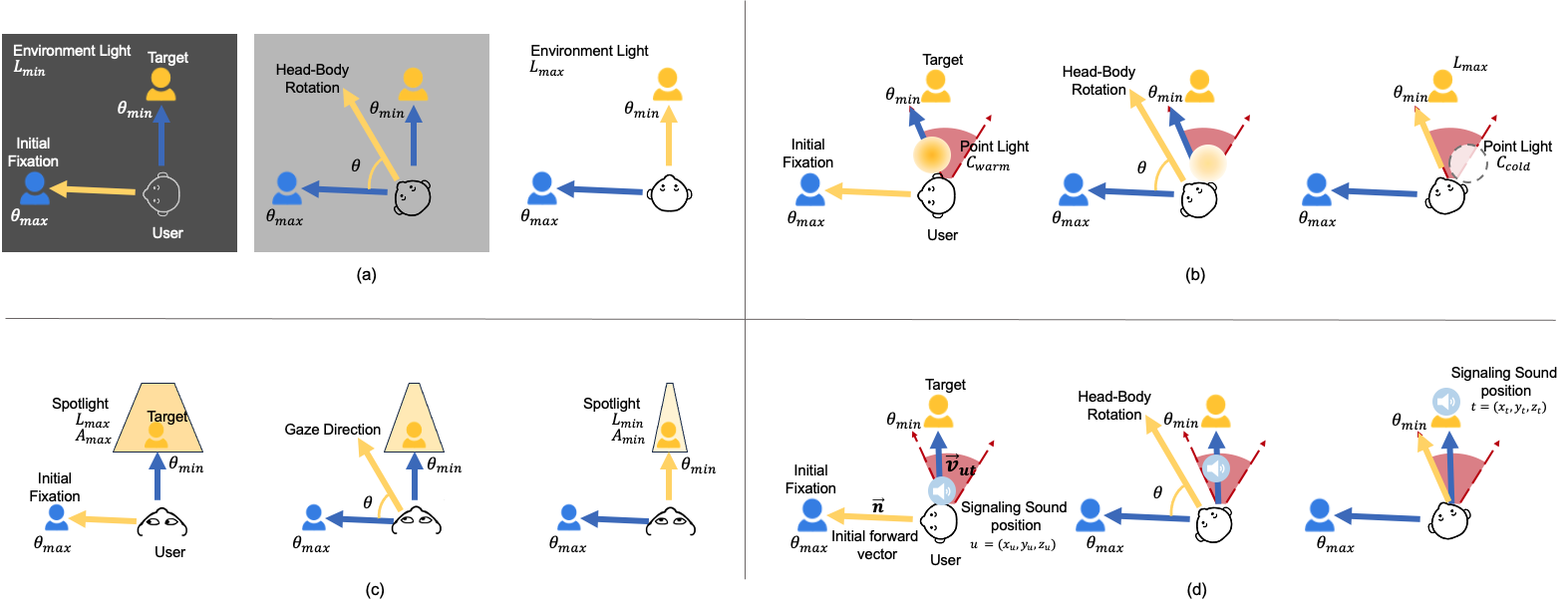}
        \vspace{-1.5em}
	\caption{Progressive adjustments in the light manipulator and spatial audio control module components relative to the angular distance ($\theta$) between the user's head-body rotation or gaze direction and the new speaker's coordinate, with subfigures (a) environment light, (b) point light, (c) spotlight, and (d) signaling sound demonstrating the range from maximum to minimum angular thresholds. The area colored in red represents $\theta_{\text{viewport}}$, the range that the new speaker coordinates is within the user's viewport.}
    \vspace{-1em}
    \label{fig:formula_figure}  
\end{figure*}

\section{Multi-modal Attention Guidance for Social VR}
\label{sec:method}
Our multi-modal attention guidance method for Social VR, informed by a literature review in group conversation psychology and insights from the formative study (Section~\ref{sec:formative_study}), integrates key factors to determine user engagement in group conversations. 

Engagement, defined as the degree of active involvement in the conversation, also reflects perceived social proximity to new speakers. For example, a participant deeply engaged in a dialogue might have a higher social distance, requiring more effort for a new speaker to attract their attention.

It is important to note that in our method, the term 'user' refers to a participant already engaged in the conversation (speaker or listener), distinct from the `new speaker' who is attempting to interject. The method employs several parameters for gauging engagement: (1) New Speaker Coordinate, representing the 3D position of a participant intending to interject, defined in the world coordinate system; (2) Head-Body Rotation, reflecting the user's head position; and (3) Gaze Direction, indicating the user's viewing direction.

These parameters form the foundation of our heuristic approach for determining the optimal level of guidance intrusiveness. Utilizing these, we developed two key modules: the \textbf{Light Manipulator Module} and the \textbf{Spatial Audio Control Module}.

\subsection{Light Manipulator Module}
Following the design characteristics drawn from the group interview study, we chose to adopt lighting effects as the core element of our attention guidance method. This choice was informed by several factors: (1) the ubiquitous nature of light in any environment, lending a diegetic quality to its usage, and (2) the potential for nuanced control of light effects, enabling us to finely tune the subtlety of the effect.

We manipulate three types of lighting sources in our approach: environmental light, point light, and spotlight, each strategically used for out-of-view and within-view attention guidance scenarios. This manipulation involves precise control over parameters such as intensity and color warmth, using the Unity Light object's parameters for accurate integration. We define `out-of-view' as when the target is outside the user's viewport angle and `within-view' as the opposite.

\paragraph{Environment Light:}
Environment light management involves dimming its brightness and restoring original luminance upon user acknowledgment of the new speaker, applicable in both out-of-view and within-view scenarios. This process entails setting the environment light's minimum intensity to a pre-defined parameter, based on the desired subtlety, with the unaltered brightness as the maximum intensity.

When a new speaker signals their intention to speak, the environmental light's intensity decreases over two seconds to a set minimum, ensuring a subtle, non-disruptive environmental shift for the user.

At the same time, angular ranges $[\theta_{min}, \theta_{max}]$ are determined for adaptively controlling light brightness. Here, $\theta_{max}$ represents the angular deviation from the user's gaze direction to the new speaker's coordinate at the signal's moment, and $\theta_{min}$ denotes the aligned angle. Ideally, $\theta_{min}$ would be 0, but we allow non-zero magnitude for flexibility.

Let $\theta$ be the angle between the user's current gaze direction and the gaze direction at the moment the new speaker's signal is received. The environment light brightness $L$ adjusts according to the function $ f^{env}(\theta)$, as shown below:
\vspace{-1em}
\begin{equation}
\begin{aligned}
    L = L_{\text{min}} + (L_{\text{max}} - L_{\text{min}}) \cdot \left(\frac{\min(\theta_{\text{max}}, \max(\theta, \theta_{\text{min}}))}{\theta_{\text{max}} - \theta_{\text{min}}}\right)^\gamma
\end{aligned}
\label{eq:envlight}
\end{equation}

where $\gamma$ is a curvature parameter adjusting the intensity change rate relative to $\theta$ (note that $\gamma > 0$). Figure.~\ref{fig:formula_figure}-(a) depicts how $L$ changes in accordance with $\theta$. The min and max functions in Equation~\ref{eq:envlight} ensure the brightness remains within desired limits, preventing excessive darkening or brightening outside the range $ [\theta_{min}, \theta_{max}]$. This mechanism maintains appropriate lighting even if the user's gaze deviates significantly, supporting ongoing conversation engagement.

\begin{figure}
 \centering
  \includegraphics[width=0.47\textwidth]{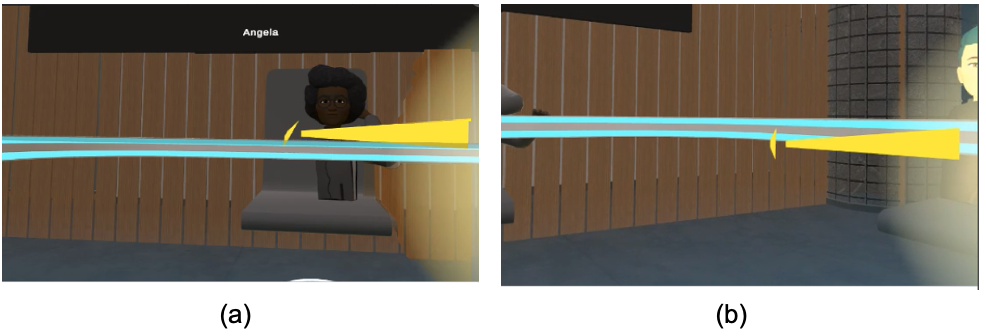}
  \caption{The point light gradually interpolates from (a) a warm color(yellow) to (b) a cold color (white) as $\theta$ gets  closer to $\theta_{min}$}
  \label{fig:pointlight}
  \vspace{-1.5em}
\end{figure}

\paragraph{Point Light:}
When the new speaker is entirely outside the user's field of view, offering a directional cue becomes imperative to guide the user's head orientation. To fulfill this purpose, we utilize a point light affixed to the user's head position, placed at $75^\circ$ to remain within peripheral vision. The point light's direction, left or right, is contingent on the new speaker's position, offering balanced guidance. We further adjust the color warmth ($C$) of the point light depending on the angular deviation ($\theta$) from the target, following the function:

\vspace{-1em}
\begin{equation}
    C = C_{\text{org}} + (C_{\text{warm}} - C_{\text{cold}}) \cdot \left(\frac{\min(\theta_{\text{max}}, \max(\theta, \theta_{\text{min}}))}{\theta_{\text{max}} - \theta_{\text{min}}}\right)^\gamma  
    \label{eq:pointlight}
\end{equation}

Influenced by the visual attention model~\cite{el2009dynamic}, which favors warm colors for attracting attention, we vary the RGB value of the point light based on the angular divergence between the user's current view and the target. This transition from a `warm' to a `cold' color is contingent on $\theta$. Note that our `warm' color ($C_{\text{warm}}$) and `cold' color ($C_{\text{cold}}$) were heuristically set as `yellow' and `white'. The point light deactivates when the new speaker coordinate is within the user's viewport. The parameter $\gamma$ controls the rate of change in color warmth with respect to $\theta$, with $\gamma = 1$ as the default setting for a linear relationship, though it can be adjusted as needed. As depicted in Figure.~\ref{fig:formula_figure}-(b), a higher $\theta$ yields a warmer point light, which decreases in warmth as the user's gaze aligns with the target. The area colored in red, $\theta_{\text{viewport}}$, represents when the new speaker coordinate is within the user's viewport, in which the point light deactivates.

\paragraph{Spotlight:}
The spotlight only activates when the target is within the user’s viewport. Hence, here $\theta$ is defined as the angular deviation of the gaze direction, not the head-body rotation. The angle of the spotlight's cone and the intensity of the brightness will dynamically adjust based on the user's gaze direction relative to the new speaker coordinate. Regarding the intensity control, the same angular distance equation as described in Equation.~\ref{eq:envlight} is applied. The cone's angle ($A$) defines the size of the area in which the spotlight covers and can be formulated as a function of $\theta$. The equation is as follows:

\begin{equation}
    A = A_{\text{min}} + (A_{\text{max}} - A_{\text{min}}) \cdot \left(\frac{\min(\theta_{\text{max}}, \max(\theta, \theta_{\text{min}}))}{\theta_{\text{max}} - \theta_{\text{min}}}\right)^\gamma
    \label{eq:spotlight}
\end{equation}

When the user's gaze is directed further away from the new speaker coordinate ($\theta_{max}$), the brightness of the spotlight increases up to its maximum value $L_{max}$, and the angle of the spotlight widens up to $A_{max}$. Conversely, when the user's gaze focuses directly on the target ($\theta_{min}$), the spotlight's brightness and range decrease $(L_{min}, A_{min})$, potentially even deactivating the spotlight. Figure.~\ref{fig:formula_figure}-(c) depicts how the parameter changes. The sensitivity at which the spotlight range ($A$) changes on varying angular deviation ($\theta$) can again be controlled through $\gamma$. The $\gamma= 1 $ will lead to a linear relation of $A$ and $\theta$.  $\gamma > 1$ will lead to a steeper decrease of $R$ as $\theta$ gets smaller. The $\theta < \gamma < 1$ will lead to a gradual decrease of $A$ as $\theta$ gets smaller. We use $\gamma = 1$ as the default but the value can be configured as needed.

\subsection{Spatial Audio Control Module}
Audio plays a crucial role in signaling and notifying users about new information. Spatial audio is particularly advantageous as it allows us to not only provide audible cues but also direct users' attention to the source of the sound. It has been shown that in hybrid video calls spatializing participants' voices was preferred to an increased speech stream identification~\cite{hyrkas2023spatialized}.

In the Spatial Audio Control Module of our system, we manage two types of sound sources: a signaling sound to indicate a user who wishes to speak and adjustments to the volume of the current speaker in the group conversation.

\paragraph{Signaling Sound:}
To notify users of a participant awaiting their turn to speak, we employ an arbitrary beeping sound akin to a chiming tone.  Based on their role as a speaker or the predetermined subtlety weight, the sound source will be projected to the user's head position. This positioning amplifies the sound and makes it appear closer to the user, indicating the need to turn their head. When the user shifts their gaze and the new target speaker enters their field of view, the sound source returns to its original position—coincident with the new speaker's coordinates. This transition of sound source coordination is proportionately controlled when it is outside the user's viewport.
Figure.~\ref{fig:formula_figure}-(d) visualizes the sound source location scenario. We denote the user’s position in 3D space as $u=(x_u,y_u,z_u)$ and the target speaker’s position as $t=(x_t,y_t,z_t)$. At the moment when the target speaker signals their intention to speak, the angular distance between the user’s initial forward vector $\vec{n}$ (i.e. where they are facing) and the vector pointing from $u$ to $t$, denoted as $\vec{v} _{ut}$ form angle of $\theta_{max}$. If we denote $o$ as the 3D position of the sound source location, it can be expressed as follows:
\begin{equation}
    o' =
\begin{cases}
    u, & \text{if } \theta \geq \theta_{\text{max}} \\
    u + | \vec{v}_{ut} | \cos(\theta_{\text{max}} - \theta), & \text{if } \theta_{\text{max}} > \theta > \theta_{\text{min}} \\
    t, & \text{if } \theta \leq \theta_{\text{min}},
\end{cases}
\label{eq:sound}
\end{equation}
where the equation ensures that the attached object is initially projected to the user’s head position. As the user’s rotation shifts from $\theta_{max}$ to $\theta_{min}$, the sound source follows the path along the vector t connecting the user and the new speaker coordinate. Finally, when $\theta$ becomes less than or equal to $\theta_{min}$, the sound source returns to the original position of the target speaker.

\paragraph{Speaker Volume:}
In addition, we lower the volume (intensity) of the current speaker for 2 seconds as the chiming sound plays. Afterward, the volume returns to its normal level. This design choice ensures that participants do not feel pressured to halt their speech. Notably, speaker volume adjustments are exclusive to listener users, as speakers do not hear their own voices through the VR headset but in their real-life surroundings. This constraint precludes us from controlling their audio levels.

\begin{figure}
  \includegraphics[width=0.5\textwidth]{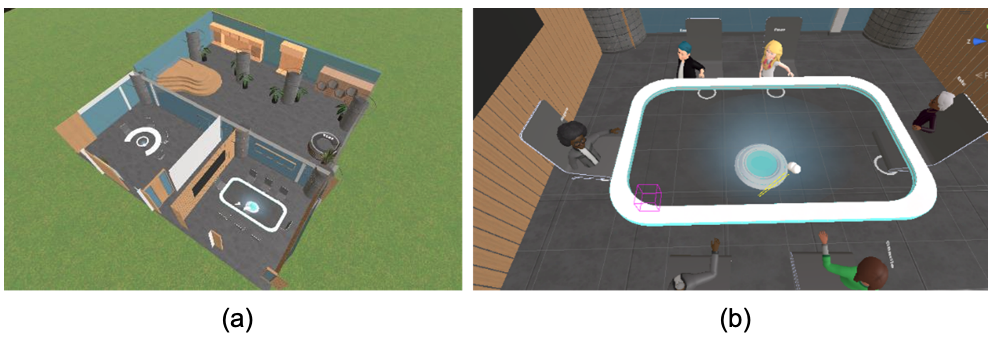}
  \vspace{-0.5em}
  \caption{We illustrate the virtual environment from (a) the top-down view and (b) with the virtual agent avatars}
  \label{fig:setup}
\end{figure}

\begin{figure*}
	\centering
	\includegraphics[width=1.0\textwidth]{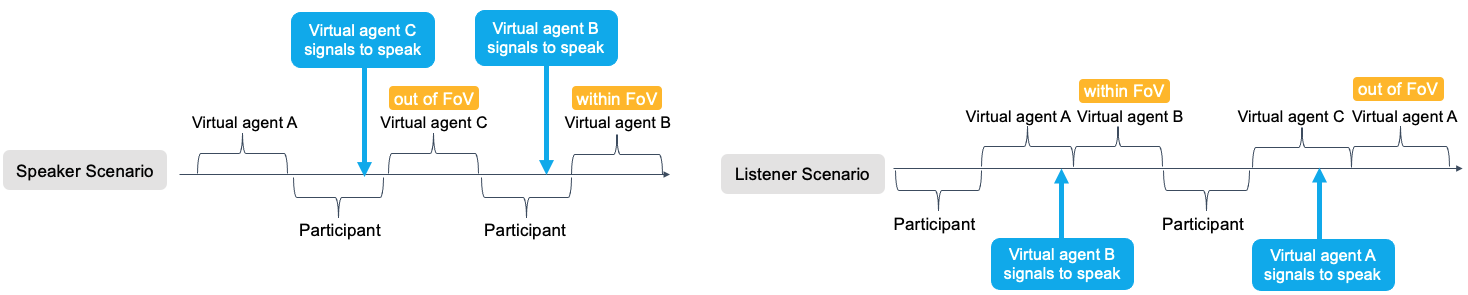}
        \vspace{-1.5em}
	\caption{Sequential turn-taking order for speaker and listener scenarios, with virtual agents' animations and voiceovers executed accordingly. The signal for the new speaker is dispatched 5 seconds following the current speaker's turn. Each scenario includes instances of the attention guidance method activation, both when the new speaker is out of and within the participant's field of view.}

    \label{fig:exp_scenario}  
\end{figure*}

\section{Evaluation Study}
\label{sec:eval}
\subsection{Study Design}
The evaluation of our approach is undertaken from the perspective of the signal \emph{receiver}, i.e., the user actively engaged in an ongoing conversation, who should be made aware of a new speaker within the group intending to contribute. This section assesses the performance of our approach using various attention guidance methods aimed at directing attention to a new speaker.

In our study, we compare two primary attention guidance methods: the \emph{Text-Icon} method, which is inspired by Horizon Workrooms—a state-of-the-art application~\cite{horizonworkrooms}. It features a text notification window displaying the user's name who is `raising their hand', accompanied by a hand icon above the new speaker's avatar. We chose this method as a baseline for comparison due to its resemblance to the ``raise-hand" feature prevalent in video call platforms, ensuring user familiarity and its current implementation in commercial VR conference systems. Additionally, we assess the \emph{SGD} method~\cite{grogorick2017subtle}, noted for its subtlety and favorable evaluation in section~\ref{sec:formative_study}, to evaluate how attention guidance methods not specifically tailored for social VR contexts are received.

\begin{figure}
  \includegraphics[width=0.5\textwidth]{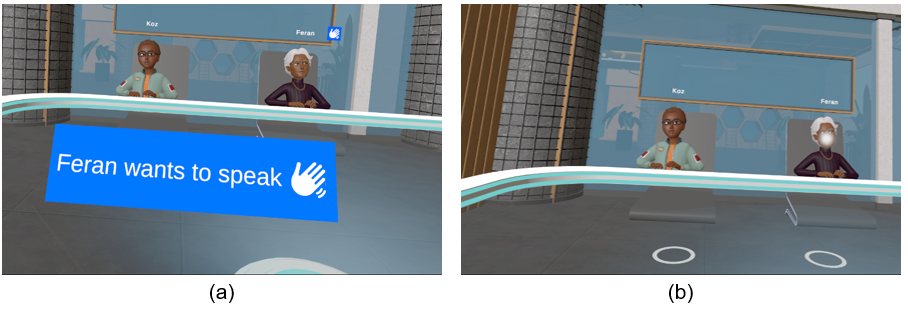}
  \caption{We compared our approach to (a) \textit{Text-Icon}, a text window fixed to the user's desk space and an icon appearing next to the agent's name tag, and (b) \textit{SGD} a flickering effect on the target at user's peripheral view.}
  \label{fig:comparison}
\end{figure}

Our investigation also includes testing our proposed method with (\emph{Light-Audio}) and without audio (\emph{Light}) to determine the effectiveness of integrating light and audio modules. The methods \emph{Text-Icon} and \emph{SGD} are depicted in Figure~\ref{fig:comparison}. Furthermore, we present a table that outlines the characteristics of each method based on a taxonomy from prior work~\cite{rothe2019guidance}, as shown in Table~\ref{tab:method_taxonomy}.

\begin{table}[ht]
\centering
\resizebox{\columnwidth}{!}{
\begin{tabular}{llll}
\hline
\textbf{Method}    & \textbf{Diegesis} & \textbf{Senses}      & \textbf{Placement}       \\ \hline
\textit{Text-Icon} & Non-diegetic  & Visual           & World-fixed          \\
\textit{SGD}               & Non-diegetic  & Visual           & Screen-fixed         \\
\textit{Light}             & Diegetic      & Visual           & World/Screen-fixed   \\
\textit{Light-Audio}       & Diegetic      & Visual/Auditory  & World/Screen-fixed   \\ \hline
\end{tabular}
}
\caption{Comparison of attention guidance methods in Section~\ref{sec:eval}}
\label{tab:method_taxonomy}
\end{table}

Our study uses a within-subjects design, ensuring the counterbalanced presentation of each method through a Latin square design. Participants utilized an Oculus Quest Pro headset, which supports eye tracking.

Considering the influence of a user's role (speaker or listener) on gaze patterns~\cite{maran2021visual}, we designed two scenarios: one where the user receives a signal while speaking and another while listening. A simulated multi-user VR meeting featured five pre-recorded virtual agents engaged in simple small talk topics. Eight distinct topics were utilized, divided evenly between the listener and speaker scenarios. Pre-recorded scripts were crafted to simulate interaction, with virtual agents asking the participant’s viewpoint on the topic and responding when the participant posed a pre-determined question to the group. It should be noted that participants were instructed beforehand on the specific question to direct to the group.

In a scenario, turn-taking occurred among the user, a virtual agent within the user's field of view, and another virtual agent positioned outside the user's field of view. The turn-taking order and timing of the new speaker's signal for each scenario are depicted in Figure~\ref{fig:exp_scenario}. To prevent user anticipation of the turn-taking order, which could lead to predictive identification of the next speaker, we employed a randomized approach. Specifically, we randomized the name and avatar graphic skinning for each virtual agent before every topic, and alternated the users' positions between two seats, ensuring an equal distribution along the topics.

\subsection{Measurements}
\label{sec:measure}

The study collected both quantitative and qualitative metrics. Quantitative measures included response time, defined as the time from signal issuance to the user aligning their gaze with the target; and a ``missed" count if the user failed to turn their head towards the target within 5 seconds. A brief quiz assessing users' attention to the conversation was also administered.

For qualitative evaluation, we administered a communication satisfaction questionnaire~\cite{hecht1978conceptualization} to investigate how the turn-taking method employed could influence the overall communication experience. The Igroup Presence Questionnaire (IPQ)~\cite{schubert2001experience} was utilized to determine the impact of our attention guidance method on users' sense of immersion. Additionally, the Notification questionnaire~\cite{ghosh2018notifivr} was employed to assess the method's effectiveness in signaling a new speaker's intervention. User preferences were quantified using a 5-point Likert scale, providing a comparative metric of user favorability across the tested methods.


\subsection{Implementation Setup}
For the virtual environment, a virtual meeting room environment was implemented with Unity version 2021.3.11f1. We utilized several SDKs for our implementation: Movement SDK for eye tracking to discern user saliency and viewing direction, Meta Avatar SDK for rendering avatars, gestures, and lip-syncing, and the Steam Audio Plugin to incorporate spatial audio features. The resulting virtual scene is shown in Figure~\ref{fig:setup}. The voiceover recordings of the avatars were generated through Speechify~\cite{speechify}. Avatar animation and lip sync were manually recorded by motion capture to ensure there were no socially odd gestures.

A preliminary study was conducted to test the functionality of the prototype and to set the values of parameters described in Section~\ref{sec:method} for the evaluation study. As a result, we empirically set the parameters as; $L_{min} = 0.5$, $L_{max} = 1.1$ for environment light; $C_{warm} = (1, 0.902, 0.259)$ (close to yellow), $C_{cold} = (1, 1, 1)$ for point in RGB color 0 to 1 scale, $L_{min} = 0.8$, $L_{max} = 1.5$ and $A_{min} = 30$, $A_{max} = 60$ for spotlight. Note that these parameters are directly applied to Unity's light object component.


\subsection{Participants}
We recruited 20 volunteers, comprising eleven females and nine males, with ages ranging from 19 to 36 years ($\mu=28.45$, $\sigma=5.25$). All participants exhibited normal color vision, with ten having unaided normal vision, and the remaining individuals possessing vision corrected to normal. Among the participants, eleven had prior experience with VR, while none had experience specifically in social VR applications.

\begin{figure*}
	\centering
	\includegraphics[width=\textwidth]{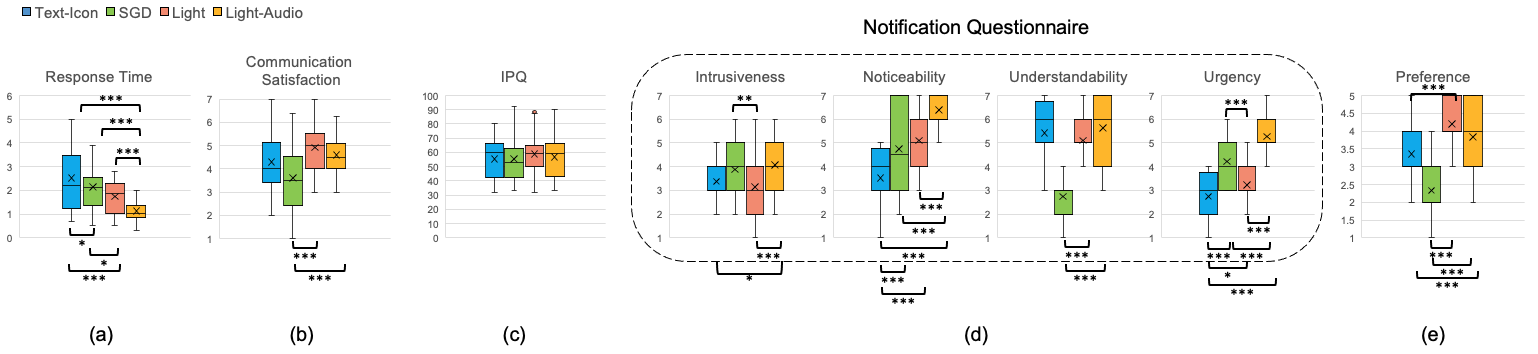}
	\caption{Statistical results for (a) response time (b) communication satisfaction (b) presence (c) preference, and (d) Notification questionnaire. The post hoc analysis revealed that \textit{Light-Audio} has a significantly faster response time, better-perceived conversation satisfaction, and preference. It is also shown that participants reported \textit{Light-Audio} to have high scores in all Notification subscales. Asterisk (*) indicates a statistically significant difference between conditions: $p < 0.05 (*); p < 0.01 (**); p < 0.001 (***)$. }
    \label{fig:exp_results}
\end{figure*}

\subsection{Procedure}
Upon welcoming the participants, we presented an overview of the study's procedures, obtained their consent through a signed form, and gathered demographic information. Seated in swivel chairs, participants were introduced to the Oculus Quest Pro and underwent eye tracker calibration. To acclimate to the virtual environment, participants briefly explored their surroundings. The nature of the study was explained, emphasizing that participants would engage in a VR conversation with a group of people without specifying that these individuals were pre-recorded avatars, aiming to enhance participant engagement. 

Each participant was assigned the username ``\textit{Charlie}" for the virtual meeting, which other virtual agents would use as a reference. Participants were instructed to respond when addressed by the agents. In the listener scenario, where the participant initiated the conversation, individuals were given a topic and instructed to commence the group discussion by posing a question to the entire group without specifying an agent to answer immediately. The conversation script was executed in a manner where the experimenter manually triggered a specific voiceover and animation at the correct timing, adhering to the predetermined turn-taking order, of which participants remained unaware. The signal, or the presentation of the method, was dispatched 5 seconds after the current speaker spoke, ensuring that participants spoke for at least 5 seconds guaranteed sufficient time to transmit the signal when the participant was the speaker. 
If the participant's gaze direction aligned with the new speaker and this alignment lasted for over 1.5 seconds, the subsequent speaker's animation would be triggered. If the previous speaker's animation was still playing, it would be interrupted. Alternatively, should the participant fail to notice the new speaker within 5 seconds, the new speaker's animation would commence.

Each trial's conversation spanned two to three minutes, and participants underwent one trial per method, resulting in four trials for each speaker/listener scenario and eight trials in total with distinct conversation topics. The entire experiment lasted approximately 70 minutes. After each trial, participants were required to fill out the questionnaires detailed in Section~\ref{sec:measure}. After the entire experiment, a semi-structured interview was conducted to gather additional comments and qualitative feedback. The experimenter regularly checked if participants needed a break, allowing breaks as necessary.

\begin{table*}[ht]
    \centering
    \begin{adjustbox}{max width=\textwidth}
        \begin{tabular}{|c|*{20}{c|}}
            \hline
            & \multicolumn{5}{c|}{Response Time} & \multicolumn{5}{c|}{Conversation Quality} & \multicolumn{5}{c|}{Presence} & \multicolumn{5}{c|}{Preference}\\
            \hline
            Factor & \(df_{\text{effect}}\) & \(MS\) & \(F\) & \(p\) & Partial \(\eta^2\) & 
            \(df_{\text{effect}}\) & \(MS\) & \(F\) & \(p\) & Partial \(\eta^2\) & 
            \(df_{\text{effect}}\) & \(MS\) & \(F\) & \(p\) & Partial \(\eta^2\)  & 
            \(df_{\text{effect}}\) & \(MS\) & \(F\) & \(p\) & Partial \(\eta^2\) \\
            \hline
            M & 3 & 28.013 & 41.954 & $< .001$ & .293 &
                3 & 12.578 & 9.545 & $< .001$ & 0.159 & 
                3 & 105.966 & .473 & .701 & .009 & 
                3 & 25.342 & 29.607 & $< .001$ & .369\\
            V & 1 & 28.301 & 42.385 & $< .001$ & .122&
                - & - & - & - & - & 
                - & - & - & - & - & 
                - & - & - & - & -\\
            R & 1 & 9.041 & 13.540 & $< .001$ & .043 & 
                1 & .018 & .014 & .906 & .000 & 
                1 & .165 & .001 &  .978 & 0.000 & 
                1 & .100 & .117 & .733 & .001 \\
            M$\times$V & 3 & 14.975 & 22.427 & $< .001$ & .181 & 
                 - & - & - & - & - & 
                - & - & - & - & - & 
                - & - & - & - & - \\
            M$\times$R & 3 & .860 & 1.288 & .279 & .013 & 
                3 & .023 & .018 & .997 & .000 & 
                3 & 15.860 & .071 & .975 & .001 & 
                3 & .183 & .214 & .886 & .004 \\
            V$\times$R & 1 & 1.986 & 2.974 & 0.086 & 0.010 & 
            - & - & - & - & - & 
            - & - & - & - & - & 
            - & - & - & - & - \\
            M$\times$V$\times$R & 3 & .161 & .242 & .867 & .002 & 
            - & - & - & - & - & 
            - & - & - & - & - & 
            - & - & - & - & - \\
    \midrule
    \addlinespace
    \midrule
            & \multicolumn{5}{c|}{Intrusiveness} & \multicolumn{5}{c|}{Noticeability} & \multicolumn{5}{c|}{Understandability} & \multicolumn{5}{c|}{Urgency}\\
            \hline
            Factor & \(df_{\text{effect}}\) & \(MS\) & \(F\) & \(p\) & Partial \(\eta^2\) & 
            \(df_{\text{effect}}\) & \(MS\) & \(F\) & \(p\) & Partial \(\eta^2\) & 
            \(df_{\text{effect}}\) & \(MS\) & \(F\) & \(p\) & Partial \(\eta^2\)  & 
            \(df_{\text{effect}}\) & \(MS\) & \(F\) & \(p\) & Partial \(\eta^2\)\\
            \hline
            M & 3 & 7.373 & 5.042 & .002 & 0.91 &
                3 & 55.940 & 32.410 & $<.001$ & .937 & 
                3 & 71.217 & 30.146 & $<.001$ & .375 & 
                3 & 50.273 & 49.928 & $<.001$ & .496\\
            R & 1 & 6.006 & 4.108 & .044 & .026 & 
                1 & .156 & .091 & .764 & .001 & 
                1 & .000 & .000 &  1.000 & .000 & 
                1 & .056 & .056 & .813 & .000 \\
            M$\times$R & 3 & .456 & .312 & .817 & .006 & 
                3 & .056 & .033 & .992 & .001 & 
                3 & .117 & .050 & .985 & .001 & 
                3 & .106 & .106 & .957 & .002 \\
            \hline
        \end{tabular}
    \end{adjustbox}
    \caption{ANOVA main effects and interactions for the Notification questionnaire (M: \textit{Method},  V: \textit{View}, R: \textit{Role}). Note that \textit{View} were not evaluated for the subjective questionnaires as a trial encompasses both within-view and out-of-view cases.}
    \label{tab:statistical_results}
\end{table*}

\subsection{Results}
We conducted a statistical analysis of our results. It is important to note that all participants correctly answered all three questions in the post-trial quiz, indicating their attentiveness to the conversational task. Additionally, no presented methods were missed by any participant in any trial. This may be attributed to the brevity of the conversation and the limited number of method presentations in each trial.

In our analysis, we consider the impact of several factors:
the presence or absence of the target within the participant's view (denoted as \textit{View}),
the role of the participant (speaker or listener, denoted as \textit{Role}), 
and the presented method (denoted as \textit{Method}).
We follow the analytical approach outlined by Rzayev et al.~\cite{rzayev2019notification}, which applied the Aligned Rank Transform (ART) using the ARTool toolkit and applied paired-sample t-test with Tukey correction and for ANOVAs used paired-sample t-test with Bonferroni correction. We depict statistical results in this section in Figure.~\ref{fig:exp_results}.


\paragraph{\textbf{Response Time:}}
There were statistically significant main effect for all \emph{Method}, \emph{View}, and \emph{Role}. A two-way interaction effect of \emph{Method}$\times$\emph{View} was observed. The pairwise comparison of \emph{Method} revealed that all comparisons were statistically significant to each other (\(p < .001\), except for between \textit{Light} and \textit{SGD} with $p = .013$ and between \textit{SGD} and \textit{Text-Icon} with $p = .017$)

Pairwise comparisons for \emph{View} revealed that response times within view were statistically significantly shorter than those out of view (\(p < .001\)). Similarly, for \emph{Role}, listener scenarios exhibited statistically significantly shorter response times compared to speaker scenarios (\(p < .001\)).

\textit{Light-Audio} led to a significantly faster response time than other methods for within-view new speakers (with \textit{Light-Audio} \(p = .001\), with \textit{SGD} \(p < .001\), and with \textit{Text-Icon} \(p = .010\)) and out-of-view new speakers (all \(p < .001\)). Compared to other methods, \textit{Text-Icon} resulted in the least response time for out-of-view signals (all \(p < .001\)). See Figure.~\ref{fig:exp_results}-(a).

\paragraph{\textbf{Communication Satisfaction}}

We found a statistically significant main effect of \textit{Method} on participants' communication experience. Pairwise comparisons showed that participants had significantly lower communication satisfaction with \textit{SGD} compared to \textit{Light} ($p < .001$) and \textit{Light-Audio} ($p < .001$).

\paragraph{\textbf{Presence:}}
While \textit{Light-Audio} and \textit{Light} showed a higher IPQ score than \textit{Text-Icon} and \textit{SGD}, there were no statistically significant main and interaction effects found in the statistical analysis (\(p > .05\)). 

\paragraph{\textbf{Notification:}}
The Notification questionnaire consists of four scales: Intrusiveness, Noticeability, Understandability, and Urgency. 
Statistical analysis revealed a significant main effect of \emph{Method} across all scales. Notably, Intrusiveness demonstrated both a significant main effect of \emph{Role}. No significant interaction effect was observed.

In terms of Intrusiveness, pairwise comparisons indicated that the speaker role yielded significantly higher intrusiveness scores than the listener role (\(p = .044\)). Regarding \emph{Method}, significant differences were observed between \textit{Light-Audio} and \textit{Light} (\(p < .001\)), \textit{Light-Audio} and \textit{Text-Icon} (\(p = .011\)), and \textit{Light} and \textit{SGD} (\(p = .008\)).

For Noticeability, the pairwise comparison showed that  \textit{Light-Audio} was significantly more noticeable than all other methods (\(p < .001\)) while \textit{Text-Icon} was significantly less noticeable than all other methods (\(p < .001\)).

For Understandability, the pairwise comparison revealed that \textit{SGD} was significantly less `understandable' than all other methods ($p < .001$). No other pair showed significant differences.

For Urgency, the pairwise comparison revealed statistically significant differences between 
all methods, with \textit{Light-Audio} having the highest score (\(p < .001\) for all pairs except for \emph{Light} and \emph{Text-Icon}, $p = .036$).
See Figure.~\ref{fig:exp_results}-(e).

\paragraph{\textbf{Preference:}}
There was a statistically significant main effect of \textit{Method} on participants' preference scores. No other statistically significant main or interaction effects were found. Pairwise comparisons revealed significant differences ($p < .001$) in preference scores between \textit{Light-Audio} and \textit{SGD}, \textit{Light} and \textit{SGD}, \textit{Light-Audio} and \textit{Text-Icon}, and \textit{Light} and \textit{Text-Icon}. See Figure.~\ref{fig:exp_results}-(d).

\section{Discussion}
\label{sec:discussion}
In this section, we explore the results and discuss possible design implications derived from them.

\subsection{Effect of Audio Cues in Our Approach}
The results analysis unveiled intriguing effects of the Audio module in our approach. The outcomes suggest that the audio cue is effective in alerting users and conveying information, as evidenced by significantly faster response times and higher noticeability. The fact that it scored significantly higher in Urgency implies the potential to dynamically add or detach the audio module based on the urgency of user conversations. This could be an interesting avenue for further evaluation, especially considering the perspective of new speakers. However, the significantly high perceived intrusiveness suggests that there may be cases where users dislike the intrusive nature of audio cues. Nonetheless, it is promising that users rated a high preference for this method.

Expanding on this, it would be valuable to explore user preferences concerning the customization of the audio module. For instance, investigating whether users would prefer adjustable volume levels or the ability to choose specific sounds could contribute to a more tailored and user-friendly experience.

\subsection{Perceived Interruption in Conversational Flow}
Interestingly, we observed an effect related to whether the participant was a speaker or a listener when receiving the new speaker signal. Participants generally perceived a better conversational experience when they were listeners compared to when they were the speakers. A participant noted, ``\textit{I felt slightly interrupted when there was a cue while I was talking. I wasn’t completely bothered by some of the methods, but I guess it’s natural to feel that way.}'' This is consistent with findings from previous research indicating that users are more likely to visually engage with others when they are listeners in a group conversation, as opposed to when they are the speakers~\cite{maran2021visual}.

Six participants verbally expressed lower satisfaction with the \textit{SGD} method compared to others, citing interruptions to their conversational flow as the primary reason for its reduced scoring. This aligns with the higher perceived intrusiveness scores of \textit{SGD}. Extending this observation, further investigation into user reactions during different stages of conversation, such as during critical points or pauses, could provide deeper insights into how these cues impact the natural flow of communication.


\subsection{Preferred Subtlety in Attention Guidance}
Participants highlighted reasons for favoring \textit{Light}, emphasizing its ability to blend into the environment. A participant stated, ``\textit{I liked that the lighting effect provided just enough cues to make me notice that there is a change in the environment, but also I can easily ignore the fact and revisit the direction the signal came from whenever I want without feeling interrupted in between.}'' While users expressed a preference for \textit{Light-Audio}, concerns were raised regarding real-world applications. A participant commented, ``\textit{I was fine with the chiming sound here since this was a short small talk among strangers, but I will be quite annoyed by the frequency of the chiming sound if it constantly comes up. It would be great if we can turn off the sound or the frequency of playback can be controlled.}'' Surprisingly, \textit{SGD} was disliked as users found the constant modulation effect in their peripheral vision annoying.

Building on this, exploring user-friendly controls for adjusting the subtlety of cues, such as the intensity of light or the frequency of audio, could enhance the applicability of these methods in various real-world scenarios.

Aligning with our group interview study, users consistently expressed a preference for characteristics of subtlety and diegeticness in attention guidance methods in a conversational setting. Further research into the nuanced aspects of these preferences could contribute to the development of more user-centric and adaptable communication systems.


\section{Conclusions, Limitations, and Future Work}
\label{sec:limitation}

We present a multimodal approach for guiding user attention in a social VR group conversation setting. Our method aims to offer tailored attention guidance, providing more pronounced cues for highly engaged participants and subtler cues for those less engaged, ultimately enhancing the overall meeting experience. Leveraging light and spatial audio as diegetic guidance methods within the virtual environment, our approach demonstrated significantly reduced response times while maintaining high perceived conversation quality and preference. Moving forward, we envision extending our work to diverse VR social scenarios, including presenter-audience relationships and dynamic party-like settings with multiple small groups forming and disbanding continuously.

Although our work has achieved notable results, it has some limitations. 
Firstly, in our formative study, interviewees were presented with existing methods using video forms, not experiencing them within VR themselves, due to computational constraints among the interviewees. This approach suggests that different findings might emerge if the methods were presented in an immersive VR environment. 
In our evaluation study, a pre-recorded setup was employed, necessitated by the requirement for multiple actors to represent different avatars in each scenario. This approach was taken to avoid participant familiarization with specific virtual agents and to maintain consistency. Additionally, it mitigated potential network connectivity issues that could adversely affect the conversational experience. However, it is important to note that communication satisfaction scores might vary in real-life settings and may present more nuanced outcomes. In future work, we plan to incorporate real-user conversations using the Desert Survival Task (DST~\cite{laferty1928desert}), as employed in prior studies~\cite{li2022conversation}.

The brevity and informal nature of the conversations conducted in this study may also have impacted the results. Additionally, the brevity and informality of the conversations. Further investigation in more formal or presentation-like settings, where participants engage in longer and structured discourse, could yield diverse outcomes.

Furthermore, our virtual environment's default lighting setups, optimized for typical scenarios, may not suit environments with different lighting conditions, such as dark rooms. A promising avenue for future work involves exploring user perceptions of `subtle' and `intrusive' lighting in varied settings and defining parameter thresholds accordingly. Another potential research direction is the integration of automatic engagement detection, utilizing non-verbal cues like nodding or physical gestures. This would provide a more comprehensive understanding of user engagement dynamics in virtual social interactions.

As we progress, it is essential to consider the challenge of managing multiple participants who wish to speak using our approach. The visualization and signaling of the order of users and queuing mechanisms demand careful consideration. Future work should delve into how to effectively represent and manage user queues, ensuring a smooth and organized communication flow.

\bibliographystyle{abbrv-doi-hyperref}

\bibliography{eba}

\end{document}